\documentclass[prl,twocolumn,floatfix]{revtex4}
\usepackage{graphicx}
\usepackage{amsmath}

\newcommand{\Up}{\uparrow}
\newcommand{\Do}{\downarrow}

\begin{document}

\title{Spin-Flip Noise in a Multi-Terminal Spin-Valve}
\author{W. Belzig}
\affiliation{Departement f\"ur Physik und Astronomie, Klingelbergstr. 82, 4056
  Basel, Switzerland}
\author{M. Zareyan}
\affiliation{Max-Planck-Institut f\"ur Physik komplexer Systeme,  
  N\"othnitzer Str. 38, 01187 Dresden, Germany}
\affiliation{Institute for Advanced Studies in Basic
  Sciences, 45195-159, Zanjan, Iran} 
\date{\today}

\begin{abstract}
We study shot noise and cross correlations in a four terminal
spin-valve geometry using a Boltzmann-Langevin approach. The Fano
factor (shot noise to current ratio) depends on the magnetic
configuration of the leads and the spin-flip processes in the normal
metal. In a four-terminal geometry, spin-flip processes are
particular prominent in the cross correlations between terminals
with opposite magnetization.
\end{abstract}



\maketitle

The discovery of the giant magneto resistance effect in magnetic
multi-layers has boosted the interest in spin-dependent transport in
the last years (for a review see e.g. \cite{bauer:97}). In combination
with quantum transport effects the field is termed spintronics
\cite{prinz:98}.  In recent experiments spin-dependent transport in
metallic multi-terminal structures has also been demonstrated
\cite{jedema:00}. One important aspect of quantum transport is the
generation of shot noise in mesoscopic conductors
\cite{blanter:00,nazarov:03}. Probabilistic scattering in combination
with Fermionic statistics leads to a suppression of the shot noise
from its classical value \cite{khlus:87,lesovik:89,buettiker:90}.


A particular interesting phenomenon are the nonlocal correlations
between currents in different terminals of a multi-terminal structure.
For a non-interacting fermionic system the cross correlations are
generally negative \cite{buettiker:92}. In a one-channel beam splitter
the negative sign was confirmed experimentally
\cite{henny:99,oliver:99}. If the electrons are injected from a
superconductor, the cross correlations may change sign and
become positive \cite{martin:96,boerlin:02,samuelsson:02,taddei:02}.
In these studies, however, the spin was only implicitly present due to
the singlet pairing in the superconductor.

Current noise in ferromagnetic - normal metal structures, in which the
spin degree of freedom plays an essential role, has so far attracted
only little attention. Non-collinear two-terminal spin valves have
been studied in \cite{brataas:02} and it was shown that the noise
depends on the relative magnetization angle in a different way than
the conductance. Thus, the noise reveals additional information on the
internal spin-dynamics. Noise has been exploited to study the
properties of localized spins by means of electron spin
resonance\cite{engel:00}. Quantum entanglement of itinerant spins can
also be probed through noise measurements \cite{loss:00}.

\begin{figure}[tbp]
  \centering
  \includegraphics[width=0.9\columnwidth]{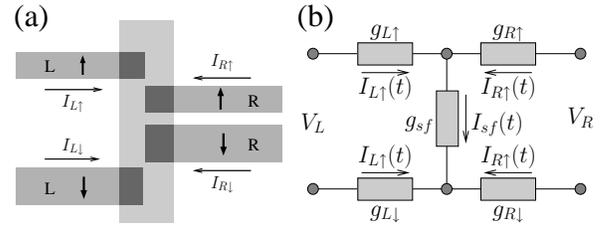}
  \caption{Four-terminal setup to measure spin-flip correlations. (a)
    a possible experimental realization with a normal diffusive metal
    strip, on which four ferromagnetic strips are deposited (of
    different width to facilitate different magnetization
    orientations). The total length of the diffusive metal underneath
    the ferromagnetic contacts should be less that the spin-diffusion
    length in the normal metal. (b) theoretical model of the device.
    Spin $\Up(\Do)$ current is flowing in the upper(lower) branch.
    Spin-flip scattering connects the two spin-branches and is modelled
    as resistor with also induces additional flutuation.}
  \label{fig:fourterm}
\end{figure}

In this work we propose a new instrument for the study of
spin-dependent transport: the use of cross correlations in a
multi-terminal structure.  The basic idea is to use a four-terminal
structure like sketched in Fig.~\ref{fig:fourterm}. An electron
current flows from the left terminals to the right terminals and is
passing a scattering region. In the absence of spin-flip scattering
the currents of spin-up electrons and spin-down electron are
independent, and the cross correlations between any of the two
currents in different spin channels vanish. However, spin-flip
scattering can convert spin-up into spin-down electrons and vice
versa, and induces correlations between the different spin currents.
This has two effects.  First, the equilibration of the
spin-populations leads to a weakened magneto-resistance effect.
Second, the current cross correlations between the differently
polarized terminals contain now information on the spin-flip processes
taking place in the scattering region.

To this end we will study a four-terminal structure, in which the
currents can be measured in all four terminals independently. The
layout is shown in Fig.~\ref{fig:fourterm}, in which the various
currents are defined.  For simplicity, we assume that all four
terminals are coupled by tunnel junctions to one node. The node is
assumed to have negligible resistance, but provides spin-flip
scattering.  The ferromagnetic character of the terminals is modelled
by spin-dependent conductances of the tunnel junctions.  The two
left(right) terminals have chemical potential $V_1(V_2)$. In most of
the final results we will assume zero temperature, but this is not
crucial. Furthermore, we will assume fully polarized tunnel contacts,
characterized by $g_{a\sigma}$, where $a=L,R$ denotes left and right
terminals, and $\sigma=\Up,\Do$ stands for the spin directions (in
equations we take $\Up=+1$ and $\Do=-1$).

The current fluctuations in our structure can be described by a
Boltzmann-Langevin formalism \cite{nagaev:92}. The time-dependent currents
at energy $E$ through contact $a\sigma$ are written as
\begin{eqnarray}
  \label{eq:i}\nonumber
  I_{a\sigma}(t,E) & = & g_{a\sigma}
  \left[f_{a\sigma}(E)-f_{c\sigma}(E)-\delta f_{c\sigma}(t,E)\right]\\
  &&+\delta I_{a\sigma}(t,E)\,.
\end{eqnarray}
The averaged occupations of the terminals are denoted by
$f_{a\sigma}(E)$, the one of the central node by $f_{c\sigma}(E)$. The
occupation of the central node is fluctuating as $\delta
f_{c\sigma}(t,E)$. The Langevin source $\delta I_{a\sigma}(t,E)$
induces fluctuations due to the probabilistic scattering in contact
$a\sigma$. We assume elastic transport in the following, so all
equations are understood to be at the same energy E. Since we assume
tunnel contacts, the fluctuations are Poissonian and given by
\cite{blanter:00}
\begin{eqnarray}
  \label{eq:tunnel-fluct}
  &&\langle \delta I_{a\sigma}(t)\delta
  I_{a^\prime\sigma^\prime}(t^\prime) \rangle
  =\\\nonumber&& g_{a\sigma}
  \delta_{\sigma\sigma^\prime}\delta_{aa^\prime}\delta(t-t^\prime) 
  \left[f_{a\sigma}+f_{c\sigma}-2f_{a\sigma}f_{c\sigma}\right]\,.
\end{eqnarray}
The brackets $\langle\cdots\rangle$ denote averaging over the
fluctuations. The conservation of the total current at all times $t$
leads to the conservation law \cite{displace}
\begin{eqnarray}
  \label{eq:itot-conservation}
  \sum_{a,\sigma}I_{a\sigma}(t)=0
\end{eqnarray}
The equation presented so far describe the transport of two
unconnected circuits for spin-up and spin-down electrons, i.e. the spin current
is conserved in addition to the total current.
Spin-flip scattering on the dot leads to a non-conserved spin current,
which we write as
\begin{eqnarray}
  \label{eq:ispin-conservation}
  \sum_{a,\sigma}\sigma I_{a\sigma}(t) & = & 
  2g_{sf}\left[f_{c\Up}+\delta f_{c\Up}(t) -f_{c\Do}-\delta
  f_{c\Do}(t)\right]\nonumber\\&&+2\delta I_{sf}(t)\,.
\end{eqnarray}
Here we introduced a phenomenological spin-flip conductance $g_{sf}$,
which connects the two spin occupations on the node. Correspondingly,
we added an additional Langevin source $\delta I_{sf}(t)$, which is
related to the probabilistic spin scattering and has a correlation
function \cite{malek}
\begin{eqnarray}
  \label{eq:sf-fluct}
  \langle \delta I_{sf}(t)\delta
  I_{sf}(t^\prime) \rangle
   & = &  g_{sf}\delta(t-t^\prime) 
  \\\nonumber &&
  \times\left[ f_{c\Up}(1-f_{c\Do})+f_{c\Do}(1-f_{c\Up})\right]\,.
\end{eqnarray}
Eqs. (\ref{eq:i})-(\ref{eq:sf-fluct}) form a complete set and
determine the average currents and the current noise of our
system. Solving for the average occupations of the node we obtain
\begin{eqnarray}
  \label{eq:fav}
  f_{c\sigma}& = & \left[\left(g_{-\sigma} g_{L\sigma}+g_{sf}g_L
    \right) f_L\right.
  \\\nonumber&&
    +\left.\left(g_{-\sigma} g_{R\sigma} + g_{sf} g_{R}\right)f_{R}\right]/Z\,.
\end{eqnarray}
Here we introduced $g_\sigma=g_{L\sigma}+g_{R\sigma}$,
$g_{L(R)}=g_{L(R)\Up}+g_{L(R)\Do}$, and $Z=g_\Up g_\Do+(g_\Up+g_\Do)g_{sf}$. 
The average currents are then
\begin{equation}
  \label{eq:currav}
  I_{L\sigma} = \frac{g_{L\sigma}}{Z}
  \left[g_{R\sigma} g_{-\sigma} + g_{R} g_{sf}\right]
  (f_L-f_R)\,,
\end{equation}
and the currents through the right terminals are obtained by
interchanging $R\leftrightarrow L$ in Eq.~(\ref{eq:currav}).
The fluctuating occupations on the node are 
\begin{eqnarray}
  \delta f_{c\sigma}(t) & = & 
  \left[ \left(g_{-\sigma}+g_{sf}\right) \delta I_\sigma(t)\right.\\\nonumber&&
    +\left.g_{sf} \delta I_{-\sigma}(t)+ g_{-\sigma} \sigma \delta I_{sf}(t) \right]/Z\,,
\end{eqnarray}
where we introduced $\delta I_\sigma(t)=\delta I_{1\sigma}(t)+\delta
I_{-1\sigma}(t)$. The total fluctuations of the current in a terminal
are obtained from $\Delta I_{a\sigma}(t)=\delta
I_{a\sigma}(t)-g_{a\sigma}\delta f_{c\sigma}(t)$ and we find
\begin{eqnarray}
  \label{eq:currfluc}
  \Delta I_{L\sigma} & = & 
  \frac{1}{Z}\left[\left(g_{R\sigma}g_{-\sigma}+\left(g_{-\sigma}+g_{R\sigma}\right)g_{sf}\right)
    \delta I_{L\sigma}
  \right.\nonumber
  \\ &&
  \nonumber
    -g_{L\sigma}\left(g_{-\sigma}+g_{sf}\right)\delta I_{R\sigma}
    +\sigma g_{L\sigma}g_{-\sigma}\delta I_{sf}
    \\ &&
    \left.     -g_{L\sigma}g_{sf}\left(\delta I_{L-\sigma}+\delta I_{R-\sigma}\right)
    \right] \,.
\end{eqnarray}
Now we can calculate all possible current correlators in the left
terminals, defined by
\begin{eqnarray}
  \label{eq:corr}
  S_{L\sigma\sigma^\prime}&=&\int_{-\infty}^{\infty} d\tau \langle \Delta
  I_{L\sigma}(t+\tau) \Delta I_{L\sigma^\prime}(t)\rangle\,.
\end{eqnarray}
The total current noise in the left terminals is
\begin{eqnarray}
  S_L=S_{L\Up\Up}+S_{L\Do\Do}+2S_{L\Up\Do}\,.
\end{eqnarray}
Of course the same quantities can be calculated for the right
terminals. From particle conservation it follows that $S_L=S_R$, but
in the presence of spin-flip scattering the individual correlators can
differ. For convenience we also define a Fano factor $F=S_L/e|I|$,
where $I=I_{L\Up}+I_{L\Do}$ is the total current.

We will discuss general results below, but first concentrate on simple
limiting cases.  We will restrict ourselves to zero temperature from
now on.  Assuming a bias voltage $V$ is applied between the right and
the left terminals, the occupations are $f_L=1$ and $f_R=0$ in the
energy range $0\le E \le eV$. The full current noise can be written as
\begin{eqnarray}
  \label{eq:fullnoise}
  S_{L} & = & \frac{\left|eV\right|}{Z^3}\sum_{\sigma=\Up\Do} \left[
  g_{L\sigma} (g_{sf}g_{R}+g_{-\sigma}g_{R\sigma})^3 \right.\\&&\nonumber
  +g_{R\sigma}(g_{sf}g_{L}+g_{-\sigma}g_{L\sigma})^3\\&&\nonumber
  +\frac{g_{sf}}{Z}(g_{\Do}g_{L\Up}-g_{\Up}g_{L\Do})^2
  (g_{sf}g_{R}+g_{\sigma}g_{R-\sigma})\\&&\nonumber\left.
  \times(g_{sf}g_{L}+g_{-\sigma}g_{L\sigma})\right]\,.
\end{eqnarray}
For the cross correlations at the left side we find
\begin{eqnarray}
  \label{eq:crosscorr}
  S_{L\Up\Do} &= & -g_{sf}\left|eV\right|\frac{g_{L\Up}g_{L\Do}}{Z^3}
  \sum_{\sigma=\Up\Do} 
  \\&&\nonumber
  \big\{\left[g_{-\sigma}g_{R\sigma}+(g_{-\sigma}+g_{R\sigma})g_{sf}\right]
    (g_{sf}g_{R}+g_{-\sigma}g_{R\sigma})
  \\&&\nonumber
  -g_{R\sigma}(g_{-\sigma}+g_{sf})(g_{-\sigma}g_{L\sigma}+g_{sf}g_{L})
  \\&&\nonumber
  +\frac{g_{\Do}g_{\Up}}{Z}(g_{-\sigma}g_{L\sigma}+g_{sf}g_{L})
    (g_{\sigma}g_{R-\sigma}+g_{sf}g_{R})\big\}\,.
\end{eqnarray}
It can be shown, that the cross correlations are always negative, as
it should be\cite{buettiker:92}.

In the case of a two-terminal geometry two different configurations
are possible. Either both terminals have the same spin-direction, or
the opposite configuration. In the first case we can take $g_\Do=0$.
There is no effect of the spin-flip scattering and we obtain for
the Fano factor $F=(g_L^2+g_R^2)/(g_L+g_R)^2$, in agreement with the
known results \cite{blanter:00}. If the two terminals have different
spin orientations ('antiferromagnetic' configuration), the situation
is completely different, since transport is allowed only by spin-flip
scattering. We take $g_{L\Do}=g_{R\Up}=0$. The Fano factor is
\begin{equation}
  \label{eq:twotermaf}
  F=1-2g_{sf}g_Lg_R 
  \frac{(g_L+g_R)(g_L+g_{sf})(g_R+g_{sf})}{(g_Lg_R+(g_L+g_R)g_{sf})^3}\,, 
\end{equation}
where we have used the result for the mean current
$I=g_{sf}g_Lg_R/(g_Lg_R+(g_L+g_R)g_{sf})$. The Fano factor, given in
Eq.~(\ref{eq:twotermaf}) interpolates between the Poisson limit $F=1$
for $g_{sf}\ll g_L+g_R$ and the result for the double barrier junction
$F=(g_L^2+g_R^2)/(g_L+g_R)^2$ for $g_{sf}\gg g_L+g_R$, coinciding with
two-terminal 'ferromagnetic' configuration \cite{mish-prep}.

Let us now turn the four-terminal structure and study the effect of
spin-flip scattering on the spin cross correlation in lowest order in
$g_{sf}/(g_\Up+g_\Do)$.  The zero-frequency cross-correlation between
the currents in the left terminals gives
\begin{equation}
  \label{eq:ccpert}
  \frac{S_{L\Up\Do}}{|eV|} = -2g_{sf}
  \frac{g_{L\Up}g_{L\Do}}{g_\Up^2g_\Do^2}
  \left[g_{R\Up}g_{R\Do}+\frac{(g_{L\Do}g_{R\Up}-g_{L\Up}g_{R\Do})^2}{g_\Up
  g_\Do}\right]\,.
\end{equation}
The first term is also present in a spin-symmetric situation,
and is caused by the additional current path opened by the
spin-flip scattering.  The second term in the
Eq.~(\ref{eq:ccpert}) depends on the amount of spin accumulation on
the central metal, i.e. is proportional to $(f_{c\Up}-f_{c\Do})^2$.

We first consider the symmetric 'ferromagnetic' configuration
$g_{L\Up}=g_{R\Up}=g_\Up/2$ and $g_{L\Do}=g_{R\Do}=g_\Do/2$. Note, that
also $g_{L}=g_{R}$ follows in this configuration. The Fano factor of
the full current noise is $F=1/2$, i.~e. we recover the usual
suppression of the shot noise characteristic for a symmetric double
barrier structure. There is no spin accumulation in this
configuration, and, consequently, no effect of the spin-flip
scattering on the Fano factor. The cross correlations in the
'ferromagnetic' configuration are
\begin{eqnarray}
  S_{L\Up\Do} & = & -\frac{g_{sf}}{8}
  \frac{g_\Up g_\Do}{g_\Up g_\Do+ g_{sf}(g_\Up+g_\Do)}|eV|\,.
\end{eqnarray}
Thus, in the limit of strong spin-flip scattering the cross
correlations become independent on $g_{sf}$.

\begin{figure}[tbp]
  \centering
  \includegraphics[width=0.8\columnwidth,keepaspectratio,clip]{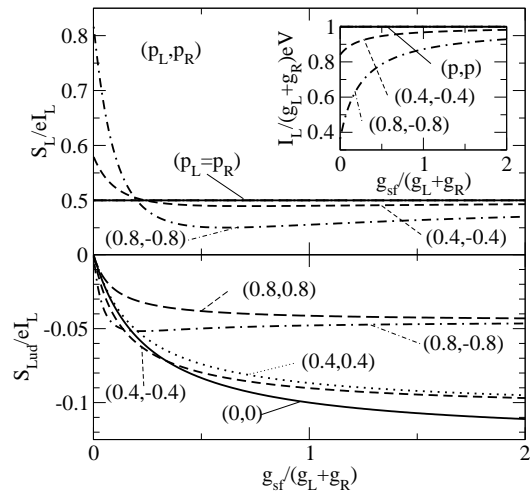}\\
  \caption{Cross correlations, Fano factor and average currents
    (symmetric case). We assume symmetric contacts $g_L=g_R$ and
    parametrize the magnetic properties with the spin polarization
    $p_{L(R)}= (g_{L(R)\Up}-g_{L(R)\Do}) / (g_{L(R)\Up}+g_{L(R)\Do})$.
    The upper part shows the Fano factor of the current fluctuations
    in the left contacts for different polarization configurations.
    Inset: average current. The lower part shows the spin-flip induced
    cross correlations between $\Up$- and $\Do$-currents in the left
    terminals.}
  \label{fig:noise1}
\end{figure}

Next we consider the symmetric 'antiferromagnetic' configuration
$g_{L\Up}=g_{R\Do}=g_1$ and $g_{L\Do}=g_{R\Up}=g_2$. The Fano factor
is
\begin{equation}
  \label{eq:af-fano}
  F=\frac{1}{2}\left[1-\frac{(g_{1}-g_{2})^2}{(g+2g_{sf})^2}
    \left(\frac{ 2g_{sf}^2}{gg_{sf}+2g_{1}g_{2}}-\frac{g}{g+2g_{sf}}\right)\right]\,.
\end{equation}
The second term in the brackets in Eq.~(\ref{eq:af-fano}) can be
either positive or negative. In the latter case $F$ drops below the
symmetric double barrier value of 1/2.  For the cross
correlations we obtain
\begin{eqnarray}
  \label{eq:af-cross}
  \nonumber
  \frac{S_{L\Up\Do}}{|eV|} & = &  
  -\frac{g_{sf}g_1g_2}{2g^2(g+2g_{sf})^4}\left[
    g(g+2g_{sf})^3\right.
  \\ & & 
  \left.+(g_1-g_2)^2\left(3g^2+6gg_{sf}+4g_{sf^2}\right)\right]\,,
\end{eqnarray}
where we introduced the abbreviation $g=g_1+g_2$. Again, the second
term in the brackets in Eq.~(\ref{eq:af-cross}) is proportional to the
spin accumulation of the island, which enhances the spin-flip induced
cross correlations.

The transport properties for symmetric junctions are shown in
Fig.~\ref{fig:noise1}. For equal polarizations of both sides there is
no effect of spin-flip scattering on the Fano factor and average
currents. However, the cross correlations do depend on the
polarizations even in this case. For small $g_{sf}$ the cross
correlations rapidly increase in magnitude.  For $g_{sf}\gg g_L+g_R$
the cross correlations become independent of the relative
polarizations. Their absolute value, however, depends strongly on the
absolute value of the polarization. For antiparallel polarizations the
Fano factor differs strongly from its value 1/2 in the unpolarized
case. With increasing spin-flip scattering rate, the Fano factor goes
from a value larger than 1/2 through a minimum, which is always lower
that 1/2.

\begin{figure}[tbp]
  \centering
  \includegraphics[width=0.8\columnwidth,keepaspectratio,clip]{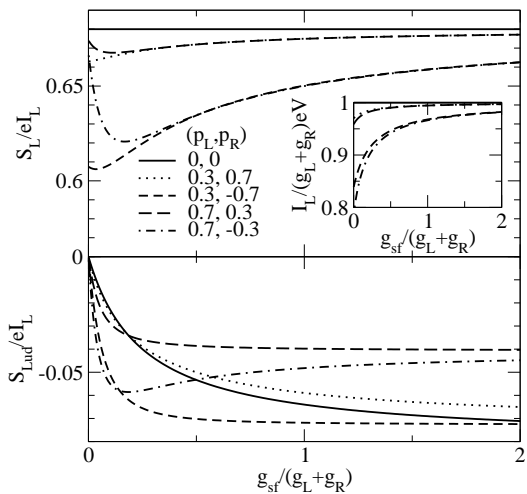}
  \caption{Cross correlations, Fano factor and average currents
    (asymmetric case). We take here $g_L=4 g_R$. The definition of the
    polarizations are taken over from Fig.~\ref{fig:noise1}.}
  \label{fig:noise2}
\end{figure}

Let us now turn to the general case of asymmetric junctions. The noise
correlations are plotted in Fig.~\ref{fig:noise2}. We have taken
$g_L=4 g_R$ and various configurations of the polarizations $0.3$ and
$0.7$.  The Fano factors and the average currents are now different
for all parameter combinations. However, the variations of the Fano
factors are small, \textit{i.~e.} they are alway close to the
unpolarized case. This is different for the cross correlations. Even
for weak spin-flip scattering they change dramatically if some of the
polarizations are reversed.



In conclusion we have suggested to use shot noise and cross
correlations as a tool to study spin-flip scattering in mesoscopic
spin-valves \cite{fourterm}. In a two-terminal device with antiferromagnetically
oriented electrodes spin-flip scattering leads to a transition from
full Poissonian shot noise (Fano factor $F=1$) to a double-barrier
behaviour ($F=1/2$) with increasing spin-flip rate. We have proposed
to measure the spin correlations induced by spin-flips in a
four-terminal device. If the spin-flip scattering rate is small, the
cross-correlation beween currents in terminal with opposite
spin-orientation gives direct access to the spin-flip scattering rate.
Presently, we have assumed fully polarized terminals, but a
generalization to arbitrary polarizations is straightforward.

We acknowledge discussion with C. Bruder. W.~B. was financially
supported by the Swiss NSF and the NCCR Nanoscience. M.~Z. thanks the
University of Basel for hospitality. During preparation of this
manuscript, a work appeared, in which a similar model was studied
\cite{sanchezprep}.


\begin{thebibliography}{99}
\bibitem{fourterm} The four-terminal structure suggested in
  Fig.~\ref{fig:fourterm}(a) is particularly suitable. Without
  applying an external magnetic field the magnetic configuration can
  be switched by changing the potentials of the different terminals.
\end{thebibliography}
\end{document}